\documentclass[showpacs,preprintnumbers,amssymb,amsmath]{revtex4}

\usepackage{graphicx}
\usepackage{dcolumn}
\usepackage{bm}

\begin{document}

\title{The $K^-\pi^+$ S-wave from the $D^+ \rightarrow K^-\pi^+\pi^+$ decay}

\affiliation{University of California, Davis, CA 95616}
\affiliation{Centro Brasileiro de Pesquisas F\'\i sicas, Rio de Janeiro, RJ, Brasil}
\affiliation{CINVESTAV, 07000 M\'exico City, DF, Mexico}
\affiliation{University of Colorado, Boulder, CO 80309}
\affiliation{Fermi National Accelerator Laboratory, Batavia, IL 60510}
\affiliation{Laboratori Nazionali di Frascati dell'INFN, Frascati, Italy I-00044}
\affiliation{Physics Department, DCI Campus Leon, University of Guanajuato, 37150 Leon, Guanajuato, Mexico}
\affiliation{University of Illinois, Urbana-Champaign, IL 61801}
\affiliation{Indiana University, Bloomington, IN 47405}
\affiliation{Korea University, Seoul, Korea 136-701}
\affiliation{Kyungpook National University, Taegu, Korea 702-701}
\affiliation{INFN and University of Milano, Milano, Italy}
\affiliation{University of North Carolina, Asheville, NC 28804}
\affiliation{Dipartimento di Fisica Nucleare e Teorica and INFN, Pavia, Italy}
\affiliation{Pontif\'\i cia Universidade Cat\'olica, Rio de Janeiro, RJ, Brazil}
\affiliation{University of Puerto Rico, Mayaguez, PR 00681}
\affiliation{University of South Carolina, Columbia, SC 29208}
\affiliation{University of Tennessee, Knoxville, TN 37996}
\affiliation{Vanderbilt University, Nashville, TN 37235}
\affiliation{University of Wisconsin, Madison, WI 53706}
\author{J.~M.~Link}
\affiliation{University of California, Davis, CA 95616}
\author{P.~M.~Yager}
\affiliation{University of California, Davis, CA 95616}
\author{J.~C.~Anjos}
\affiliation{Centro Brasileiro de Pesquisas F\'\i sicas, Rio de Janeiro, RJ, Brazil}
\author{I.~Bediaga}
\affiliation{Centro Brasileiro de Pesquisas F\'\i sicas, Rio de Janeiro, RJ, Brazil}
\author{C.~Castromonte}
\affiliation{Centro Brasileiro de Pesquisas F\'\i sicas, Rio de Janeiro, RJ, Brazil}
\author{A.~A.~Machado}
\affiliation{Centro Brasileiro de Pesquisas F\'\i sicas, Rio de Janeiro, RJ, Brazil}
\author{J.~Magnin}
\affiliation{Centro Brasileiro de Pesquisas F\'\i sicas, Rio de Janeiro, RJ, Brazil}
\author{A.~Massafferri}
\affiliation{Centro Brasileiro de Pesquisas F\'\i sicas, Rio de Janeiro, RJ, Brazil}
\author{J.~M.~de~Miranda}
\affiliation{Centro Brasileiro de Pesquisas F\'\i sicas, Rio de Janeiro, RJ, Brazil}
\author{I.~M.~Pepe}
\affiliation{Centro Brasileiro de Pesquisas F\'\i sicas, Rio de Janeiro, RJ, Brazil}
% ...
\collaboration{The FOCUS Collaboration}
\affiliation{See \textrm{http://www-focus.fnal.gov/authors.html} for additional author information.}

\date{\today}

\begin{abstract}
Using data from FOCUS (E831) experiment at Fermilab, we present a model independent
partial-wave analysis of the $K^-\pi^+$ S-wave amplitude from the decay 
$D^+ \rightarrow K^-\pi^+\pi^+$. The S-wave is a generic complex function to be 
determined directly from the data fit. The P- and D-waves are parameterized by a sum
of Breit-Wigner amplitudes. The measurement of the S-wave amplitude covers the whole 
elastic range of the $K^-\pi^+$ system.

\end{abstract}


\maketitle

\section{Introduction}

Over forty years have passed since the birth of the Constituent Quark Model, yet the
scalar mesons still challenge theoreticians and experimentalists. Many states have been reported.
Some still need confirmation, others need to have better measurements of the pole position
and couplings to specific channels.
The identification of the nature of each state --- regular $q\overline{q}$ mesons,
tetraquarks, molecules, glueballs --- is a major task which will only be 
accomplished combining results from different types of data.

An important problem is the understanding of the low energy part of the
S-wave $K^-\pi^+$ spectrum,
where the existence of an $I=1/2$ state, the $\kappa(800)$ meson, has been the subject of a 
long-standing debate. Evidence for a neutral low mass scalar state in heavy flavor decays
has been reported by several experiments \cite{cg,bes,kmat,CLEO-c}. The pole position has been
determined recently using Roy-Steiner representations of  $K^-\pi^+$ scattering \cite{dg}.
However, evidence for the charged partner is still scarce and conflicting \cite{babar,belle}. 

The primary source for the  $K\pi \to K\pi$ scattering has been the
data from the classic LASS experiment \cite{lass}, $K^-p \to K^-\pi^+n$. With a cut at low momentum 
transfers, the  $K^-p$ interaction is assumed to be entirely due to the one-pion-exchange mechanism. 
The incident pion is, therefore, not a real, asymptotically free particle, but a nearly on-shell virtual state. 
An additonal cut on the $\pi^+n$ mass was set to avoid baryonic intermediate states. 
The LASS analysis was performed on a sample containing 151 thousand  events. 
With this sample LASS found that the  $K\pi$ cross section is elastic up to
the $K\eta'$ threshold (1.454 GeV/c$^2$). Unfortunately, LASS data start only at
$m_{K\pi} =$ 825 MeV/c$^2$. 

Heavy flavor decays are currently the only way to access the whole elastic range of the 
$K\pi$ spectrum, starting from threshold. A golden mode for the neutral $K\pi$ system is the decay 
$D^+ \rightarrow K^-\pi^+\pi^+$, which has a largely dominant S-wave component --- a common feature of
three-body final states with identical pions.

This decay was already studied by the Fermilab FOCUS Collaboration \cite{kmat}. In our previous
work the $D^+ \rightarrow K^-\pi^+\pi^+$ Dalitz plot was analyzed with the K-matrix formalism,
which was applied for the first time in Dalitz plot analysis of $D$ decays 
in the FOCUS study of the $D^+ \to \pi^+\pi^-\pi^+$ decay \cite{3pi}. 
As a cross check, a fit with the usual isobar model was also performed.

\begin{figure}
\includegraphics[width=8.5 cm]{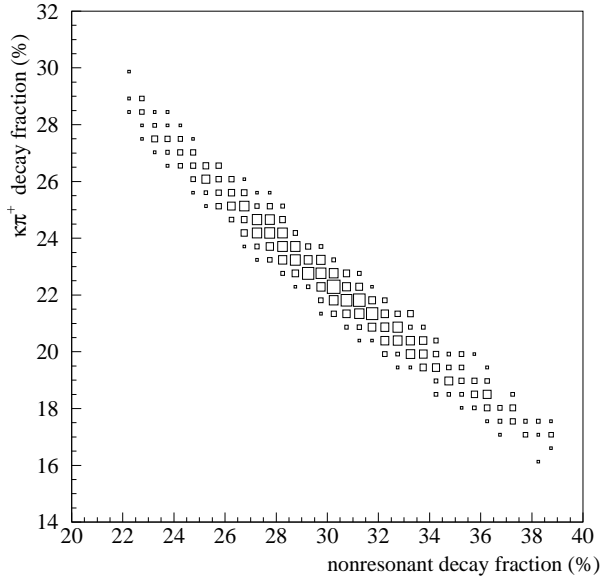}
\caption{Correlation between nonresonant and $\kappa \pi^+$ decay fractions in the isobar model.
An ensemble of  $D^+ \rightarrow K^-\pi^+\pi^+$ Dalitz plots was simulated and fitted with the 
isobar model. The scatter plot shows a high correlation between the two largest contributions of the S-wave.}
\label{corr}
\end{figure}

In the isobar model the S-wave is
represented by a coherent sum of a uniform nonresonant term plus two relativistic Breit-Wigner
amplitudes. A good fit can be achieved, but it is difficult to determine the relative 
amount of each S-wave component. 
In order to illustrate the correlation between the S-wave components of the isobar model
an ensemble of 2,000 $D^+ \to K^-\pi^+\pi^+$ Dalitz plots was simulated using the set of parameters
from our isobar fit (Table 2 of reference  \cite{kmat}). Each simulated Dalitz plot was fitted and
a scatter plot of the nonresonant versus $\kappa(800) \pi^+$ decay fractions is presented in 
Fig. \ref{corr}.
One can clearly  see that a better description of the S-wave requires one to go beyond the 
isobar model.

The K-matrix formalism is based on the assumption that there is no three-body 
final state interaction. In this approach the dynamics of the  $K^-\pi^+\pi^+$ final state are 
driven by the $K^-\pi^+$ system. Data on $D$ decays and on scattering would be directly
related and the two-body unitarity would become a constraint. The evolution of the $K^-\pi^+$ pair 
is fixed to that of $K\pi$ elastic scattering, considering the contribution of both 
$I$=1/2 and $I$=3/2 $K^-\pi^+$ amplitudes.
The parameters of the $K^-\pi^+$ production amplitude and the relative amount and phase of the 
two isospin components are determined by the fit. 
A good description of the data was obtained, with an important contribution 
of the $I=3/2$  $K\pi$ amplitude. The production amplitude has a slowly varying phase. 
The conclusion of this study is that data on $D$ decays and scattering are consistent.  The
three-body final state interactions would, therefore,  play a marginal role.

In this paper we complete our study of the $D^+ \rightarrow K^-\pi^+\pi^+$ (charge 
conjugate states are always implied) Dalitz plot, applying, to the same data set,
the model-independent partial wave analysis (MIPWA) technique, developed by the E791 Collaboration 
\cite{bm}. In this method, the $K^-\pi^+$ S-wave amplitude is parameterized by a generic 
complex function, to be determined directly from the data. The only assumption common
to all other Dalitz plot analyses is that the P- and D-waves are well represented by a sum of 
Breit-Wigner amplitudes. The $K^-\pi^+$ spectrum is divided into slices. The 
magnitude and phase of the S-wave component at the edge of each slice are determined by the fit.
A cubic spline interpolation is used to obtain the  S-wave magnitude and phase at any point in the spectrum.

The MIPWA technique provides a model-independent way to determine the $K^-\pi^+$ S-wave amplitude.
The result, however, is inclusive. The measured phase, in addition to the $I$=1/2 $K^-\pi^+$ 
phase, may contain contributions from the $I$=3/2 components, as well as possible contributions 
from three-body final state interactions.

The paper is organized as follows. In the next section we describe the selection of the data 
sample. The MIPWA formalism is described in Section III. The results of the MIPWA fit are 
presented in Section IV.

%
\vskip 0.5cm \section{The $D^+ \to K^-\pi^+\pi^+$ sample} %

FOCUS is a charm photo-production experiment  which collected data during the 1996--97 fixed target run 
at Fermilab. The photon beam was produced  by means of bremsstrahlung, from electron and positron beams
(typically with $300~\textrm{GeV}$ endpoint energy). The electron/positron beams were obtained from 
the  $800~\textrm{GeV}$ Tevatron proton beam. The photon beam interacted with a segmented BeO 
target~\cite{photon}. The mean photon energy for reconstructed charm events is $\sim 180~\textrm{GeV}$. 

The FOCUS spectrometer has a system of three multi-cell threshold \v{C}erenkov
counters to perform the charged particle identification, separating kaons from
pions up to a momentum of $60~\textrm{GeV}/c$. The identification and separation
of charm primary (production) and secondary (decay) vertecis are made by two systems of 
silicon micro-vertex detectors. The first system consists of 4 planes
of micro-strips interleaved with the experimental target~\cite{WJohns} and the
second system consists of 12 planes of micro-strips located downstream of the
target. The charged particle momentum is determined by measuring the deflections in two magnets of
opposite polarity through five stations of multi-wire proportional chambers.

The data set used in this analysis is the same as in Ref. \cite{kmat}.
The final states are selected using a \textit{candidate driven vertex
algorithm}~\cite{spectro}. A secondary vertex is formed from the three
candidate tracks. The momentum of the resultant $D^{+}$ candidate is used as
a \textit{seed} track to intersect the other reconstructed tracks and to 
search for a primary vertex. The primary vertex must have least two reconstructed tracks 
in addition to the $D^+$ seed. The confidence level of each vertex is
required to be greater than 1\%. Once the production and decay vertecis are 
determined, the distance $L$ between the vertecis and its error $\sigma _{L}$ are computed. 
The quantity $L$\thinspace /\thinspace $\sigma _{L}$ is an unbiased measure of the 
significance of detachment between the primary and secondary vertecis. This is the most 
important criterium for separating charm events from non-charm prompt backgrounds. 
Signal quality is further enhanced by isolation requirements. 
Tracks forming the $D$ candidate vertex must have a confidence level smaller than 0.001\% 
to form a vertex with the tracks from the primary vertex.
In addition, all remaining tracks not assigned to either the 
primary or the secondary vertex must have a confidence level smaller than 0.1\% to form a vertex 
with the $D$ candidate daughters. 


Particle identification cuts used in FOCUS are based on likelihood ratios between the 
various particle identification hypotheses. These likelihoods are computed for a given track
from the observed firing response (on or off) of all the cells that are
within the track's ($\beta =1$) \v{C}erenkov cone for each of our three 
\v{C}erenkov counters. The product of all firing probabilities for all the cells
within the three \v{C}erenkov cones produces a $\chi ^{2}$-like variable 
$W_{i}=-2\ln (\mathrm{Likelihood})$ where $i$ ranges over the electron, pion,
kaon and proton hypotheses~\cite{cerenkov}. The kaon track is required
to have $\Delta _{K}=W_{\pi }-W_{K}$ greater than $3$; both pion candidates
are required to satisfy $\Delta _{\pi}=W_{K }-W_{\pi}$ greater than $3$; in addition,
all tracks are required to be separated by less than $5$ units from the best
hypothesis, that is $\Delta W=W_{min}-W_{K}<5$ and $\Delta W=W_{min}-W_{\pi }<5$ .
These \v{C}erenkov cuts reduce the contamination of $D^+_s\to K^-K^+\pi^+$ background
to a negligible level.
 
\begin{figure}
\includegraphics[width=8.5 cm]{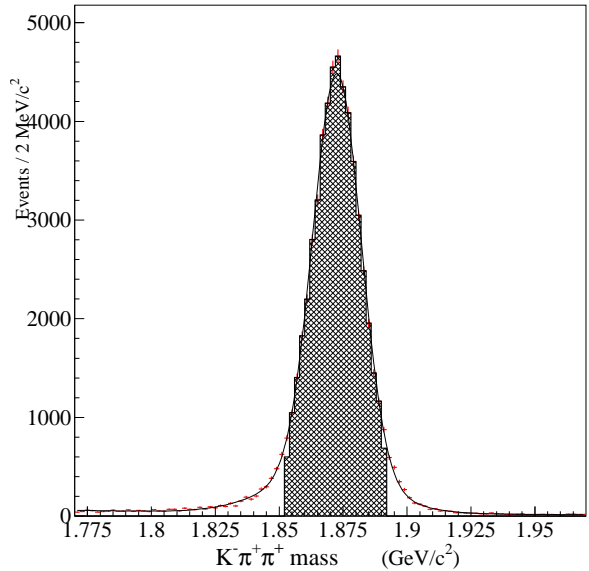}
\caption{The $K^{-}\pi^{+}\pi^{+}$ invariant mass distribution.}
\label{kpp}
\end{figure}

Using the set of selection cuts just described, we obtain the invariant $K^{-}\pi^{+}\pi^{+}$ 
mass distribution  shown in Fig. \ref{kpp}. The mass plot of Fig. \ref{kpp} is fitted with a 
function that includes two Gaussian functions with different  widths and the same mean, which take into 
account differences in the resolution in the momentum determination of our spectrometer~\cite{spectro}, and an 
exponential function for the background. The events used in the  MIPWA fit correspond to the
shaded area in Fig. \ref{kpp}, i.e., events with 1.8515 $< M_{K\pi\pi}<$ 1.9031 GeV/c$^2$. Events
in this mass region that lie outside the kinematic limit defined by the nominal $D^+$ mass
are discarded. The final data subset contains 53,595 events, with a purity (S/(S+B)) of 98.8\%.

\begin{figure}
\includegraphics[width=8.5 cm]{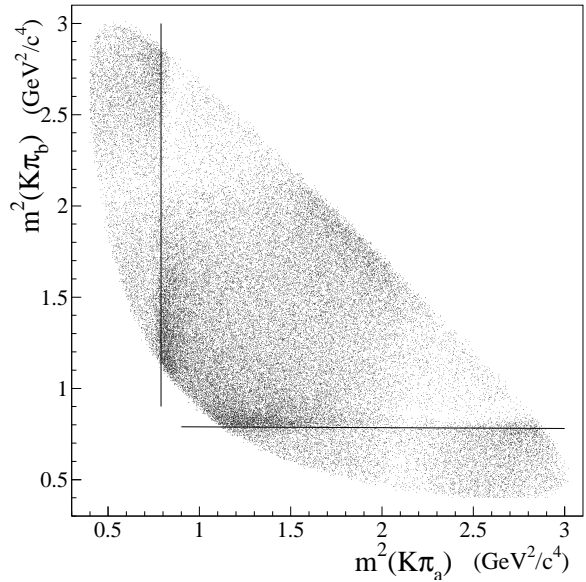}
\caption{The $K^{-}\pi^{+}\pi^{+}$ Dalitz plot.}
\label{dp}
\end{figure}

The symmetrized Dalitz plot 
of these events (two entries per event) is shown in Fig. ~\ref {dp}. A narrow band corresponding 
to the $D^+ \to \overline{K}^*(892)^0\pi^+$ events can be clearly seen. The asymmetry  in each 
$\overline{K}^*(892)^0$ lobe is evident and it is caused by the interference between this 
state and the $K^-\pi^+$ S-wave. Indeed, it is this interference with the P-wave that allows
one to access the S-wave phase.

%
\section{The Model Independent Partial Wave Analysis formalism}
%

In the MIPWA formalism
the Dalitz plot of Fig. \ref{dp} is described by a coherent sum of three partial waves,
corresponding to the $K^-\pi^+$ system in the angular momentum states $L=$0, 1 and 2.
The partial waves are complex functions of the two $K\pi$ invariant masses squared,
$s_a=(p_K+p_{\pi_a})^2$ and $s_b=(p_K+p_{\pi_b})^2$, which specify the kinematics of the
$D^+ \to K^{-}\pi^{+}_a\pi^{+}_b$ decay. Each partial wave is Bose-symmetrized with respect to 
the identical pions, 

\begin{equation}
\mathcal{A}_L = A_L(s_a,s_b) + A_L(s_b,s_a).
\end{equation}

The $K^-\pi^+$ S-wave amplitude is an unknown complex function of the $K^-\pi^+$ mass squared,

\begin{equation}
A_0(s_a,s_b) = a_0(s_a)e^{i\phi_0(s_a)} + a_0(s_b)e^{i\phi_0(s_b)}.
\label{c0}
\end{equation}

No assumption about the content of the S-wave is made:
the real functions $a_0(s)$ and $\phi_0(s)$ are determined directly by the Dalitz plot fit.
The $K^-\pi^+$ mass spectrum is divided into 39 slices of the same size. For each of the 40
endpoints  $s_k$ there are two free parameters, $a_k$ and $\phi_k$, defining the function $A_0(s_a,s_b)$ at 
that position. A cubic spline interpolation is used to define the values of both $a_0(s)$ and $\phi_0(s)$ 
between $s_k\leq s < s_{k+1}$. The S-wave has, therefore, a set of 40 pairs ($a_k,\phi_k$) of fit 
parameters.

The $K^-\pi^+$ P-wave amplitude has two components, namely the $\overline{K}^*(892)^0\pi^+$,
taken as the reference mode, and the $\overline{K}^*(1680)^0\pi^+$,

\begin{equation}
A_1(s_a,s_b) = F^D_1(s_a,s_b) F^R_1(s_a,s_b) [c_0 BW_{K^*(892)}(s_a) + c_1 BW_{K_1^*(1680)}(s_a)]\mathcal{M}_1(s_a,s_b). 
\end{equation}

The D-wave has only one component, the $\overline{K}_2^*(1430)\pi^+$ mode,

\begin{equation}
A_2(s_a,s_b) = c_2 \left[ F^D_2(s_a,s_b) F^R_2(s_a,s_b) BW_{K_2^*(1430)}(s_a)\right] \mathcal{M}_2(s_a,s_b)
\end{equation}

The complex coefficients $c_i$ are also fit parameters, except for $c_0$, the coefficient of the reference mode, which is fixed to 1.0. 

In the above equations $F^D$  and $F^R$ are the usual Blatt-Weisskopf form factors \cite{blatt},

\begin{equation}
F_{L=1} = [1+(rq)^2]^{-1/2},
\end{equation}
and
\begin{equation}
F_{L=2} = [9+3(rq)^2+(rq)^4)]^{-1/2}.
\end{equation}
where $q$ is the momentum of the resonance decay products in the resonance rest frame.
The form factor parameters 
$r=r_D$ for the D decay vertex, and $r=r_R$ for the resonance decay,
are fixed at the values used in Ref. \cite{kmat}:  $r_D=1.5$~(GeV/c)$^{-1}$ and $r_R=5.0$~(GeV/c)$^{-1}$.

The functions $\mathcal{M}_L$ are the spin amplitudes, accounting for angular
momentum conservation. For a spin-1 resonance $D^+ \to R \pi^+_b$, $R \to  K^-\pi^+_a$
the corresponding spin amplitude is

\begin{equation}
{\cal M}_{L=1} = \sum_M p_{\pi_b}^{\mu}e_{\mu}(p,M)e_{\nu}(p,M)(p_{\pi_a}-p_K)^{\nu},
\end{equation}
where $e_{\mu}$ is the resonance polarization vector with magnetic quantum number $M$, 
$p$ is the momentum 4-vector and $p_i^{\mu}$ are the momenta of the final state particles. After summing
over the unobserved resonance polarization states, the spin amplitude reduces to

\begin{equation}
{\cal M}_{L=1} =  -2\mid \vec{p}_{\pi_b} \mid \mid \vec{p}_K \mid \mathrm{cos} \ \theta,
\label{spin1}
\end{equation}
where $\theta$ is the cosine of the angle formed by $\vec{p}_K$ and $\vec{p}_{\pi_b}$ 
in the resonance rest frame.

In the case of the $\overline{K}^*_2(1430) \pi^+$ mode, the spin amplitude is

\begin{equation}
{\cal M}_{L=2} = \sum_M p_{\pi_b}^{\mu}p_{\pi_b}^{\nu}
e_{\mu}(p,M)e_{\nu}(p,M)e_{\alpha}(p,M)e_{\beta}(p,M)
(p_{\pi_a}-p_K)^{\alpha}(p_{\pi_a}-p_K)^{\beta},
\end{equation}
which, after summing over the resonance polarization, reduces to

\begin{equation}
{\cal M}_{L=2} = \frac{4}{3}(\mid \vec{p}_{\pi_b} \mid \mid
\vec{p}_K \mid)^2 (3\mathrm{cos}^2\theta-1) .
\label{spin2}
\end{equation}

The relativistic Breit-Wigner has an energy dependent width,

\begin{equation}
\mathrm{BW} =  \frac{1}{s - s_0 + i\sqrt{s_0} \ \Gamma_{\mathrm{\mathrm{tot}}}(s)},
\label{bw}
\end{equation}
where $s$ is the $K^-\pi^+$ mass squared, $s_0$ the resonance nominal mass and

\begin{equation}
\Gamma_{\mathrm{\mathrm{tot}}}(s) = \Gamma_0 
\sqrt{\frac{s_0}{s}}\left(\frac{q}{q_0}\right)^{2L+1}\frac{F_L^2(q)}{F_L^2(q_0)},
\end{equation}
where $L$ is the orbital angular momentum in the rest frame of the decaying resonance.

The signal distribution is corrected on an event-by-event basis for the acceptance. The acceptance 
includes geometry, detector and selection cuts efficiency. It is determined by a full Monte Carlo 
simulation of events: the $\gamma - N$ interaction, event propagation through the spectrometer, event 
reconstruction and selection of the $K\pi\pi$ sample. The acceptance function is obtained by  fitting the
Dalitz plot of Monte Carlo events to a 10th order polynomial.

The signal probability distribution is normalized to unity,

\begin{equation}
P_S(s_a,s_b) = \frac {1}{N_S} \varepsilon(s_a,s_b)  \left| \sum \mathcal{A}_L \right|^2,
\end{equation}
where $\varepsilon(s_a,s_b)$ is the acceptance function and $N_S$ the overall normalization constant,

\begin{equation}
N_S = \int ds_ads_b\varepsilon(s_a,s_b)  \left| \sum \mathcal{A}_L  \right|^2.
\end{equation}

The background probability distribution is fixed in the fit. The background shape is determined by 
a fit to the Dalitz plot of events from the $K^+\pi^-\pi^-$ mass sidebands \cite{kmat}. The signal 
fraction is estimated by a fit to the  $K^+\pi^-\pi^-$ mass spectrum.

In the MIPWA fit there are 40$\times$2 $+$ 2$\times$2$=$84 free parameters.
The optimum set of parameters is determined by an unbinned maximum likelihood 
fit, minimizing the quantity $w\equiv -2\ln(L)$, where the likelihood function, $L$, is given by

\begin{equation}
L = \prod_\mathrm{events} \left [f_S P_S^i(s_a,s_b) + (1-f_S) P_B^i(s_a,s_b)\right ]
\end{equation}
where $f_s$ is the signal fraction $f_s = S / (S+B)$.

Decay fractions are obtained from the coefficients $c_k$, determined by the fit,
and after integrating the overall signal amplitude over the phase space,

\begin{equation}
f_k = \frac{\int ds_ads_b\left|c_k {\mathcal A}_k(s_a,s_b) \right|^2}{\int ds_ads_b \left|
\sum_j c_j {\mathcal A}_j(s_a,s_b) \right|^2}.
\end{equation}

Errors on the fractions include errors on both magnitudes and phases, and are computed using the 
full covariance matrix.

\section{Results of the MIPWA}

The decay fractions resulting from the MIPWA fit are presented in Table \ref{pwa1}. 
For comparison, the third and fourth columns have the fractions from our previous
fits using the K-matrix formalism and the isobar 
model \footnote{The two statistical errors here reported
on the $\overline{K}^*(892)^0\pi^+$ decay fraction for both the K-matrix and isobar fits are smaller than
those quoted in Ref. \cite{kmat}. They were  overestimated in Ref. [3] as a consequence of a minor mistake in the
error propagation code, which affected their computation only and not those of the other fit fractions. The
two new values reported here are the correct ones, and should replace the old ones.}.

The data is well described by a P-wave model with two components. No improvement in the
fit quality is observed when a third component, the $\overline{K}^*(1410)\pi^+$ mode, 
is added. The contribution of this mode is consistent with zero. As in previous
analyses of the $D^+ \to K^+\pi^-\pi^-$ Dalitz plot, the S-wave component is  
dominant. The decay fractions from the MIPWA and from our previous K-matrix Dalitz plot fit are in 
good agreement.  

In Table \ref{pwa2} the MIPWA decay fractions are compared to the ones from E791 and CLEO-c.
Our results are in good agreement with the decay fractions from E791. The total S-wave contribution from CLEO-c 
is significantly higher if we add the binned and the $\overline{K}^*_0(1430)\pi^+$ fractions. 

\begin{figure}
\includegraphics[width=8.5 cm]{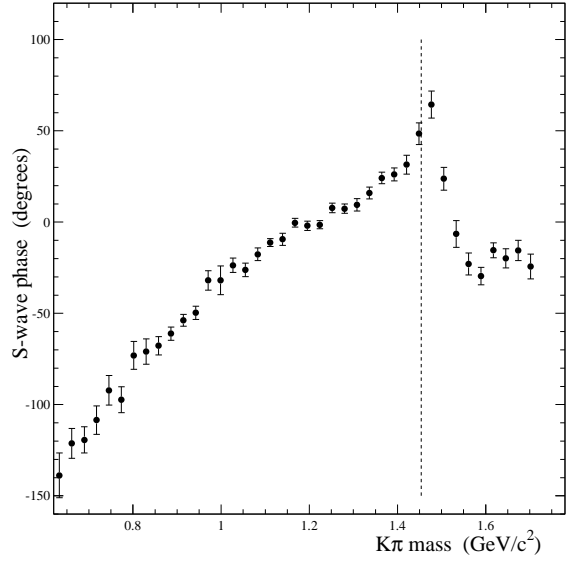}
\caption{The S-wave phase as a function of the $K^{-}\pi^{+}$ mass from the MIPWA 
$K^{-}\pi^{+}\pi^{+}$ Dalitz plot fit. The  hashed vertical line shows the elastic range 
according to LASS.}
\label{phpwa}
\end{figure}

\begin{figure}
\includegraphics[width=8.5 cm]{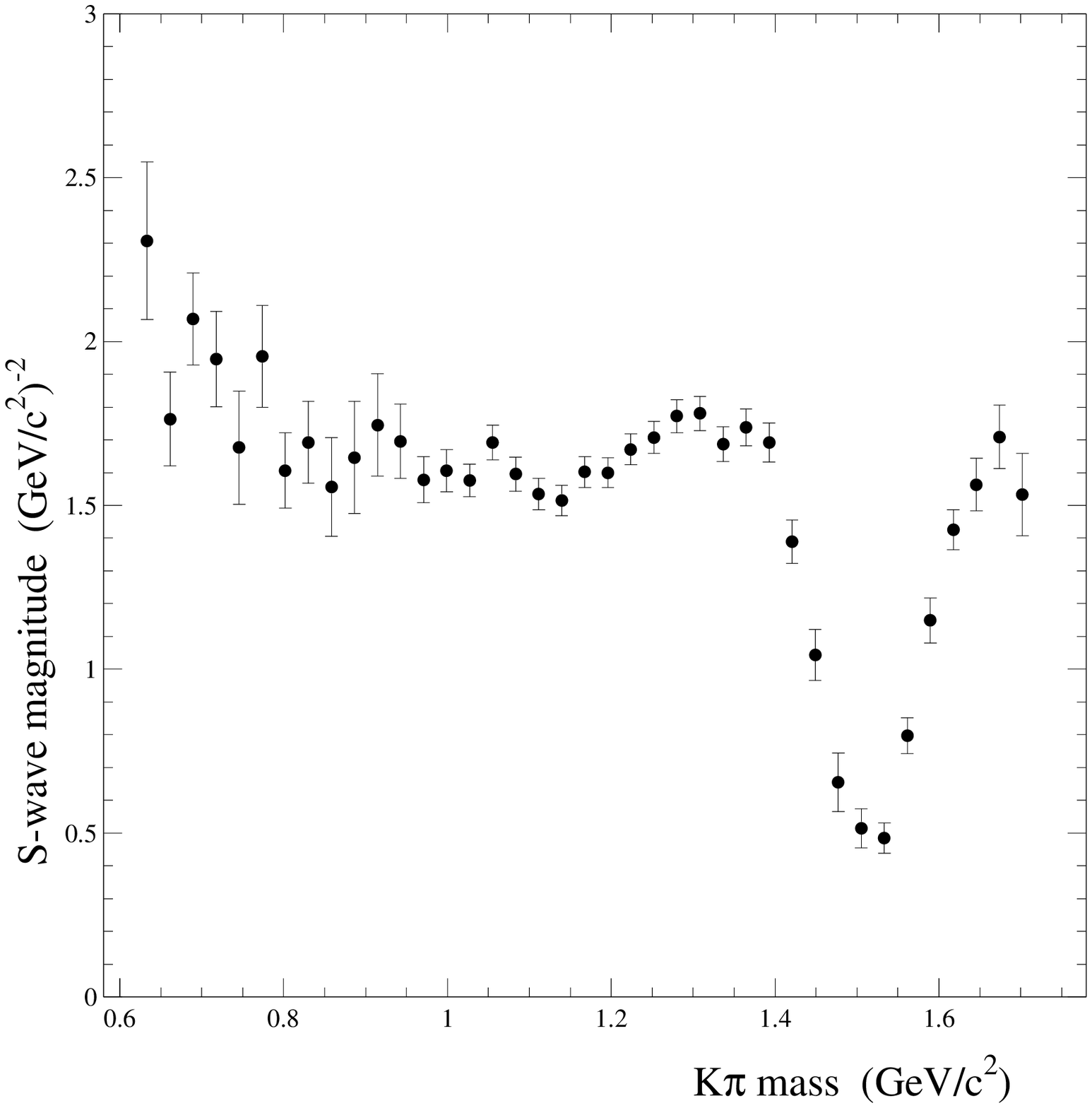}
\caption{The S-wave magnitude as a function of the $K^{-}\pi^{+}$ mass from the MIPWA 
$K^{-}\pi^{+}\pi^{+}$ Dalitz plot fit.}
\label{ampwa}
\end{figure}

The fitted values of the S-wave magnitudes and phases are presented in Table \ref{swavet} 
and plotted in Figs. \ref{phpwa} and  \ref{ampwa}. The error bars in these
figures contain the statistical and systematic errors folded in quadrature. The dashed line
in Fig. \ref{phpwa} indicates the $K\eta'$ threshold,  the upper limit of the region where 
the $K^-\pi^+$ amplitude is predominantly elastic \cite{lass}. 

The S-wave phase grows continuously across the elastic region, starting at -138$^o$ and with a total 
variation of 200$^o$. After a sudden drop near the $\overline{K}^{*}_0(1430)$ mass, 
the phase becomes nearly constant. 

The S-wave  magnitude is a decreasing function up to $m_{K\pi} \simeq$ 1.2 GeV/c$^2$. There is a dip 
near the $\overline{K}^{*}_0(1430)$ mass, which is most readily explained by the interference 
between the different components of the S-wave. 

The measured magnitudes are more affected by the systematic uncertainties than are the measured phases. In both cases
the systematic uncertainties are comparable to  or larger than the statistical errors.

\begin{table}
\caption{Decay fractions (\%) from the MIPWA Dalitz plot fit. In the MIPWA column the  first error is 
statistical, the second and third errors
are, respectively, our estimate of the {\em split sample} and {\em fit variant}  systematic uncertainties, and the last
error is the systematic error due to the uncertainty in the parameters of the other waves. Below 
the decay fractions and the phases, in degrees, are indicated.}
\begin{ruledtabular}
\label{pwa1}
\begin{tabular}{lcccccr}    
          mode                  &        FOCUS MIPWA                        &  FOCUS K-matrix                  &   FOCUS isobar model            \\
\\ \hline

$K^-\pi^+$ S-wave               & 80.24$\pm$1.38$\pm$0.23$\pm$0.25$\pm$0.26 & 83.23$\pm$1.50$\pm$0.04$\pm$0.07 &  -                              \\ 
\\
$\overline{K}^*(892)^0\pi^+$    & 12.36$\pm$0.34$\pm$0.19$\pm$0.16$\pm$0.23 & 13.61$\pm$0.41$\pm$0.01$\pm$0.30 &  13.7$\pm$0.4$\pm$0.6$\pm$0.3   \\
                                &   0 (fixed)                               &     0 (fixed)                    &	0 (fixed)	         \\
\\
$\overline{K}^*(1410)^0\pi^+$   &    -                                      & 0.48$\pm$0.21$\pm$0.012$\pm$0.17 &   0.2$\pm$0.1$\pm$0.1$\pm$0.04  \\  
                                &    -                                      & 293$\pm$17$\pm$0.4$\pm$7         &   350$\pm$34$\pm$17$\pm$15      \\
\\
$\overline{K}^*(1680)^0\pi^+$   & 1.75$\pm$0.62$\pm$0.24$\pm$0.23$\pm$0.42  & 1.90$\pm$0.65$\pm$0.01$\pm$0.43  &   1.8$\pm$0.4$\pm$0.2$\pm$0.3   \\  
                                &    67$\pm$6$\pm$2$\pm$2$\pm$3             &	 1$\pm$7$\pm$0.2$\pm$6	       &   3$\pm$7$\pm$4$\pm$8           \\
\\  
$\overline{K}^*_2(1430) \pi^+$  &  0.58$\pm$0.1$\pm$0.04$\pm$0.03$\pm$0.04  &  0.39$\pm$0.1$\pm$0.004$\pm$0.05 &  0.4$\pm$0.05$\pm$0.04$\pm$0.03 \\ 
                                &   336$\pm$7$\pm$3$\pm$2$\pm$2             &	296$\pm$7$\pm$0.3$\pm$1        &  319$\pm$8$\pm$2$\pm$2          \\
\\				
$\overline{K}^*_0(1430) \pi^+$  &  -                                        & -                                &   17.5$\pm$1.5$\pm$0.8$\pm$0.4  \\ 
                                &  -                                        & -                                &   36$\pm$5$\pm$2$\pm$1.2        \\
\\				
$\kappa \pi^+$                  &  -                                        & -                                &   22.4$\pm$3.7$\pm$1.2$\pm$1.5  \\ 
                                &  -                                        & -                                &   199$\pm$6$\pm$1$\pm$5         \\	
\\				
nonresonant                     &  -                                        & -                                &   29.7$\pm$4.5$\pm$1.5$\pm$2.1  \\ 
                                &  -                                        & -                                &   325$\pm$4$\pm$2$\pm$1.2       \\

\end{tabular}
\end{ruledtabular}
\end{table}

\begin{table}
\caption{Decay fractions (\%) and phases, in degress, from the MIPWA Dalitz plot fit compared to E791 and CLEO-c.}
\begin{ruledtabular}
\label{pwa2}
\begin{tabular}{lccccr}    
          mode                  &        FOCUS MIPWA                        &    E791          &      CLEO-c     \\
\\ \hline
$K^-\pi^+$ S-wave               & 80.24$\pm$1.38$\pm$0.23$\pm$0.25$\pm$0.26 &  78.6$\pm$2.3    &  83.8$\pm$3.8  \\ 

\\				
$\overline{K}^*_0(1430) \pi^+$  &  -                                        & -                & 13.3$\pm$0.62  \\ 
                                &  -                                        & -                &  51 (fixed)	\\				

\\
$\overline{K}^*(892)^0\pi^+$    & 12.36$\pm$0.34$\pm$0.19$\pm$0.16$\pm$0.23 & 11.9$\pm$2.0     & 9.88$\pm$0.46  \\
                                &   0 (fixed)                               &	0 (fixed)      &   0 (fixed)    \\
\\
$\overline{K}^*(1680)^0\pi^+$   & 1.75$\pm$0.62$\pm$0.24$\pm$0.23$\pm$0.42  &  1.2$\pm$1.2     & 0.20$\pm$0.12  \\  
                                &    67$\pm$6$\pm$2$\pm$2$\pm$3             & 43$\pm$17        &  113$\pm$14    \\
\\  
$\overline{K}^*_2(1430) \pi^+$  &  0.58$\pm$0.1$\pm$0.04$\pm$0.03$\pm$0.04  &  0.2$\pm$0.1     &  0.20$\pm$0.04 \\ 
                                &   336$\pm$7$\pm$3$\pm$2$\pm$2             & -12$\pm$29       &  15$\pm$9	\\

\end{tabular}
\end{ruledtabular}
\end{table}

\begin{table}
\caption{Magnitudes and phases of the S-wave from MIPWA fit. The first error is statistical. The second and third errors
are, respectively, our estimate of the {\em split sample} and {\em fit variant}  systematic uncertainties, whereas the last
error is the systematic error due to the uncertanty in the parameters of the other waves. The full systematic
error is a sum in quadrature of these three errors. This is the number between parentheses.}
\begin{ruledtabular}
\label{swavet}
\begin{tabular}{lcccr} 
$K^-\pi^+$ mass (GeV/c$^2$)& a (GeV/c$^2$)$^{-2}$ &  $\phi$ (degrees)
\\ \hline
0.63 & 2.31 $\pm$ 0.24 $\pm$ 0.02 $\pm$ 0.07 $\pm$ 0.19  ~(0.20) &  -138 $\pm$ 10 $\pm$  2 $\pm$  4 $\pm$  6  ~(7)
\\
0.66 & 1.76 $\pm$ 0.14 $\pm$ 0.07 $\pm$ 0.06 $\pm$ 0.13  ~(0.16) &  -121 $\pm$  7 $\pm$  2 $\pm$  3 $\pm$  6  ~(7)
\\
0.69 & 2.07 $\pm$ 0.14 $\pm$ 0.08 $\pm$ 0.06 $\pm$ 0.13  ~(0.16) &  -119 $\pm$  6 $\pm$  3 $\pm$  3 $\pm$  5  ~(7)
\\
0.72 & 1.95 $\pm$ 0.15 $\pm$ 0.01 $\pm$ 0.08 $\pm$ 0.14  ~(0.16) &  -108 $\pm$  5 $\pm$  2 $\pm$  3 $\pm$  6  ~(7)
\\
0.75 & 1.68 $\pm$ 0.17 $\pm$ 0.05 $\pm$ 0.09 $\pm$ 0.13  ~(0.17) &   -92 $\pm$  6 $\pm$  3 $\pm$  2 $\pm$  6  ~(7)
\\
0.77 & 1.95 $\pm$ 0.16 $\pm$ 0.01 $\pm$ 0.07 $\pm$ 0.15  ~(0.16) &   -97 $\pm$  5 $\pm$  3 $\pm$  2 $\pm$  5  ~(6)
\\
0.80 & 1.61 $\pm$ 0.11 $\pm$ 0.05 $\pm$ 0.07 $\pm$ 0.15  ~(0.16) &   -73 $\pm$  5 $\pm$  1 $\pm$  2 $\pm$  7  ~(7)
\\
0.83 & 1.69 $\pm$ 0.12 $\pm$ 0.04 $\pm$ 0.09 $\pm$ 0.15  ~(0.17) &   -70 $\pm$  4 $\pm$  4 $\pm$  1 $\pm$  4  ~(6)
\\
0.86 & 1.56 $\pm$ 0.15 $\pm$ 0.06 $\pm$ 0.08 $\pm$ 0.14  ~(0.16) &   -67 $\pm$  3 $\pm$  4 $\pm$  0 $\pm$  2  ~(7)
\\
0.89 & 1.65 $\pm$ 0.17 $\pm$ 0.03 $\pm$ 0.05 $\pm$ 0.16  ~(0.17) &   -61 $\pm$  2 $\pm$  2 $\pm$  0 $\pm$  2  ~(3)
\\
0.91 & 1.75 $\pm$ 0.16 $\pm$ 0.05 $\pm$ 0.06 $\pm$ 0.18  ~(0.19) &   -53 $\pm$  3 $\pm$  2 $\pm$  0 $\pm$  2  ~(3)
\\
0.94 & 1.70 $\pm$ 0.11 $\pm$ 0.04 $\pm$ 0.09 $\pm$ 0.15  ~(0.17) &   -49 $\pm$  4 $\pm$  2 $\pm$  0 $\pm$  2  ~(3)
\\
0.97 & 1.58 $\pm$ 0.07 $\pm$ 0.04 $\pm$ 0.05 $\pm$ 0.13  ~(0.14) &   -31 $\pm$  7 $\pm$  2 $\pm$  1 $\pm$  4  ~(5)
\\
1.00 & 1.61 $\pm$ 0.06 $\pm$ 0.03 $\pm$ 0.05 $\pm$ 0.10  ~(0.12) &   -31 $\pm$  6 $\pm$  3 $\pm$  1 $\pm$  4  ~(5)
\\
1.03 & 1.58 $\pm$ 0.05 $\pm$ 0.03 $\pm$ 0.03 $\pm$ 0.10  ~(0.11) &   -23 $\pm$  6 $\pm$  1 $\pm$  1 $\pm$  3  ~(3)
\\  
1.06 & 1.69 $\pm$ 0.05 $\pm$ 0.04 $\pm$ 0.03 $\pm$ 0.14  ~(0.15) &   -26 $\pm$  5 $\pm$  2 $\pm$  0 $\pm$  2  ~(3)
\\
1.08 & 1.60 $\pm$ 0.05 $\pm$ 0.02 $\pm$ 0.03 $\pm$ 0.12  ~(0.13) &   -17 $\pm$  4 $\pm$  2 $\pm$  0 $\pm$  2  ~(3)
\\
1.11 & 1.53 $\pm$ 0.05 $\pm$ 0.04 $\pm$ 0.02 $\pm$ 0.13  ~(0.13) &   -11 $\pm$  4 $\pm$  1 $\pm$  0 $\pm$  2  ~(2)
\\
1.14 & 1.52 $\pm$ 0.05 $\pm$ 0.03 $\pm$ 0.01 $\pm$ 0.11  ~(0.11) &    -9 $\pm$  3 $\pm$  2 $\pm$  0 $\pm$  1  ~(2)
\\
1.17 & 1.60 $\pm$ 0.05 $\pm$ 0.01 $\pm$ 0.01 $\pm$ 0.10  ~(0.10) &     0 $\pm$  3 $\pm$  1 $\pm$  0 $\pm$  1  ~(1)
\\
1.20 & 1.60 $\pm$ 0.05 $\pm$ 0.05 $\pm$ 0.01 $\pm$ 0.08  ~(0.09) &    -1 $\pm$  3 $\pm$  1 $\pm$  0 $\pm$  1  ~(1)
\\ 
1.22 & 1.67 $\pm$ 0.05 $\pm$ 0.05 $\pm$ 0.01 $\pm$ 0.07  ~(0.08) &    -1 $\pm$  3 $\pm$  1 $\pm$  1 $\pm$  1  ~(2)
\\
1.25 & 1.71 $\pm$ 0.05 $\pm$ 0.04 $\pm$ 0.01 $\pm$ 0.11  ~(0.12) &     7 $\pm$  4 $\pm$  1 $\pm$  1 $\pm$  1  ~(2)
\\
1.28 & 1.77 $\pm$ 0.05 $\pm$ 0.05 $\pm$ 0.01 $\pm$ 0.11  ~(0.12) &     7 $\pm$  4 $\pm$  1 $\pm$  1 $\pm$  1  ~(2)
\\
1.31 & 1.78 $\pm$ 0.05 $\pm$ 0.04 $\pm$ 0.02 $\pm$ 0.10  ~(0.11) &     9 $\pm$  4 $\pm$  2 $\pm$  1 $\pm$  2  ~(3)
\\
1.34 & 1.69 $\pm$ 0.05 $\pm$ 0.01 $\pm$ 0.02 $\pm$ 0.10  ~(0.10) &    15 $\pm$  4 $\pm$  1 $\pm$  1 $\pm$  2  ~(3)
\\
1.36 & 1.74 $\pm$ 0.06 $\pm$ 0.05 $\pm$ 0.02 $\pm$ 0.10  ~(0.11) &    24 $\pm$  4 $\pm$  1 $\pm$  1 $\pm$  2  ~(3)
\\
1.39 & 1.69 $\pm$ 0.06 $\pm$ 0.07 $\pm$ 0.01 $\pm$ 0.11  ~(0.13) &    26 $\pm$  5 $\pm$  1 $\pm$  1 $\pm$  2  ~(3)
\\
1.42 & 1.39 $\pm$ 0.07 $\pm$ 0.05 $\pm$ 0.01 $\pm$ 0.09  ~(0.11) &    31 $\pm$  6 $\pm$  2 $\pm$  2 $\pm$  3  ~(6)
\\
1.45 & 1.04 $\pm$ 0.08 $\pm$ 0.01 $\pm$ 0.03 $\pm$ 0.10  ~(0.10) &    48 $\pm$  6 $\pm$  3 $\pm$  3 $\pm$  4  ~(6)
\\
1.48 & 0.66 $\pm$ 0.09 $\pm$ 0.01 $\pm$ 0.05 $\pm$ 0.09  ~(0.11) &    64 $\pm$  7 $\pm$  1 $\pm$  3 $\pm$  5  ~(6)
\\
1.51 & 0.52 $\pm$ 0.06 $\pm$ 0.01 $\pm$ 0.01 $\pm$ 0.11  ~(0.11) &    23 $\pm$ 12 $\pm$  1 $\pm$  4 $\pm$  4  ~(6)
\\
1.53 & 0.48 $\pm$ 0.05 $\pm$ 0.04 $\pm$ 0.06 $\pm$ 0.08  ~(0.11) &    -6 $\pm$ 13 $\pm$  1 $\pm$  4 $\pm$  6  ~(7)
\\
1.56 & 0.80 $\pm$ 0.05 $\pm$ 0.06 $\pm$ 0.05 $\pm$ 0.09  ~(0.12) &   -23 $\pm$  9 $\pm$  1 $\pm$  2 $\pm$  5  ~(6)
\\
1.59 & 1.15 $\pm$ 0.07 $\pm$ 0.03 $\pm$ 0.08 $\pm$ 0.08  ~(0.11) &   -29 $\pm$  8 $\pm$  1 $\pm$  1 $\pm$  4  ~(4)
\\ 
1.62 & 1.43 $\pm$ 0.06 $\pm$ 0.04 $\pm$ 0.05 $\pm$ 0.09  ~(0.11) &   -15 $\pm$  7 $\pm$  1 $\pm$  2 $\pm$  3  ~(4)
\\
1.65 & 1.56 $\pm$ 0.08 $\pm$ 0.06 $\pm$ 0.09 $\pm$ 0.10  ~(0.14) &   -19 $\pm$  7 $\pm$  3 $\pm$  1 $\pm$  3  ~(4)
\\
1.67 & 1.71 $\pm$ 0.10 $\pm$ 0.02 $\pm$ 0.08 $\pm$ 0.11  ~(0.13) &   -15 $\pm$  8 $\pm$  3 $\pm$  2 $\pm$  3  ~(5)
\\
1.70 & 1.53 $\pm$ 0.13 $\pm$ 0.04 $\pm$ 0.10 $\pm$ 0.12  ~(0.16) &   -24 $\pm$  9 $\pm$  4 $\pm$  2 $\pm$  4  ~(6)
\\
1.73 & 1.60 $\pm$ 0.16 $\pm$ 0.06 $\pm$ 0.10 $\pm$ 0.15  ~(0.19) &   -34 $\pm$  14 $\pm$ 5 $\pm$  3 $\pm$  6  ~(8)

\\
\end{tabular}
\end{ruledtabular}
\end{table}

\subsection{Goodness-of-fit}

For all fits the goodness-of-fit is accessed through a two-dimensional $\chi^2$ test, using an adaptive 
binning algorithm. The folded Dalitz plot is divided into 844 cells of variable size, with a minimum occupancy of 
50 data events, in such a way that all cells have a nearly equal and sufficiently large population. This
procedure allows us to test the fit quality in great detail across the Dalitz plot.
For each cell we define the $\chi^2$ as

\begin{equation}
\chi^2_i = \frac{(n_{\mathrm{obs}} - n_{\mathrm{exp}})^2}{\sigma_{\mathrm{exp}}^2}.
\end{equation}

In the above expression $n_{\mathrm{exp}}$ is the expected population of each cell, given by a
Monte Carlo simulation performed  with 1,000,000 events generated according to the model resulting 
from the MIPWA fit, and 
$\sigma_{\mathrm{exp}}$ is the uncertainty on this number.
The overall $\chi^2$ is a sum of the $\chi^2_i$ over all cells. The number of degrees-of-freedom 
is given by the number of cells minus the number of fit parameters. From these two quantities we
estimate the confidence level of our fits.

\begin{figure}
\includegraphics[width=8.5 cm]{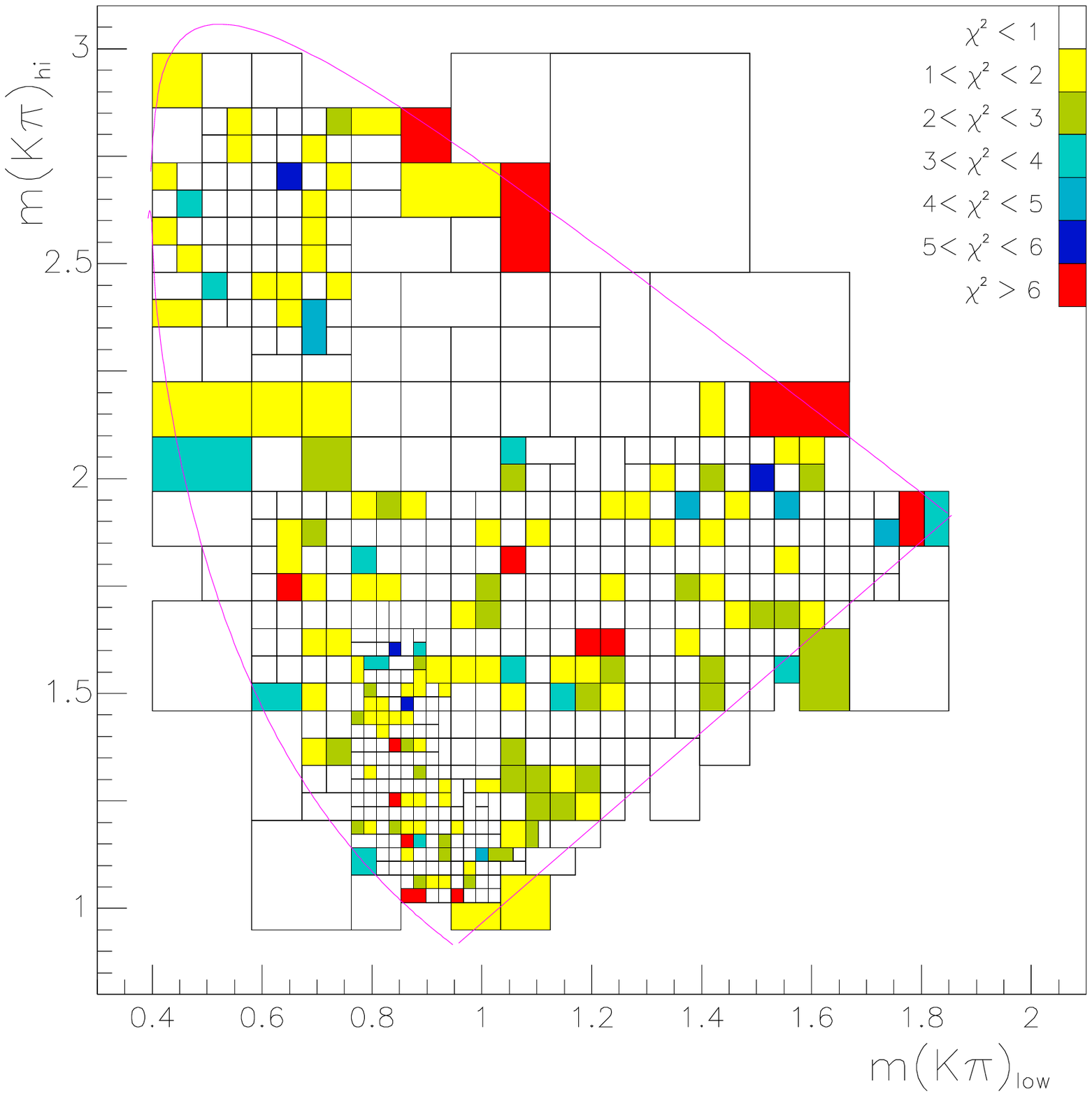}
\caption{The $\chi^2$ distribution across the folded Dalitz plot.}
\label{chi2}
\end{figure}

The overall $\chi^2$ of the MIPWA fit is $\chi^2$=818.8 (844-84=760 degrees of freedom), 
which corresponds to a confidence level of
6.8\%. The $\chi^2$ distribution across the Dalitz plot is shown in Fig. \ref{chi2}. 

\begin{figure}
\includegraphics[width=10 cm]{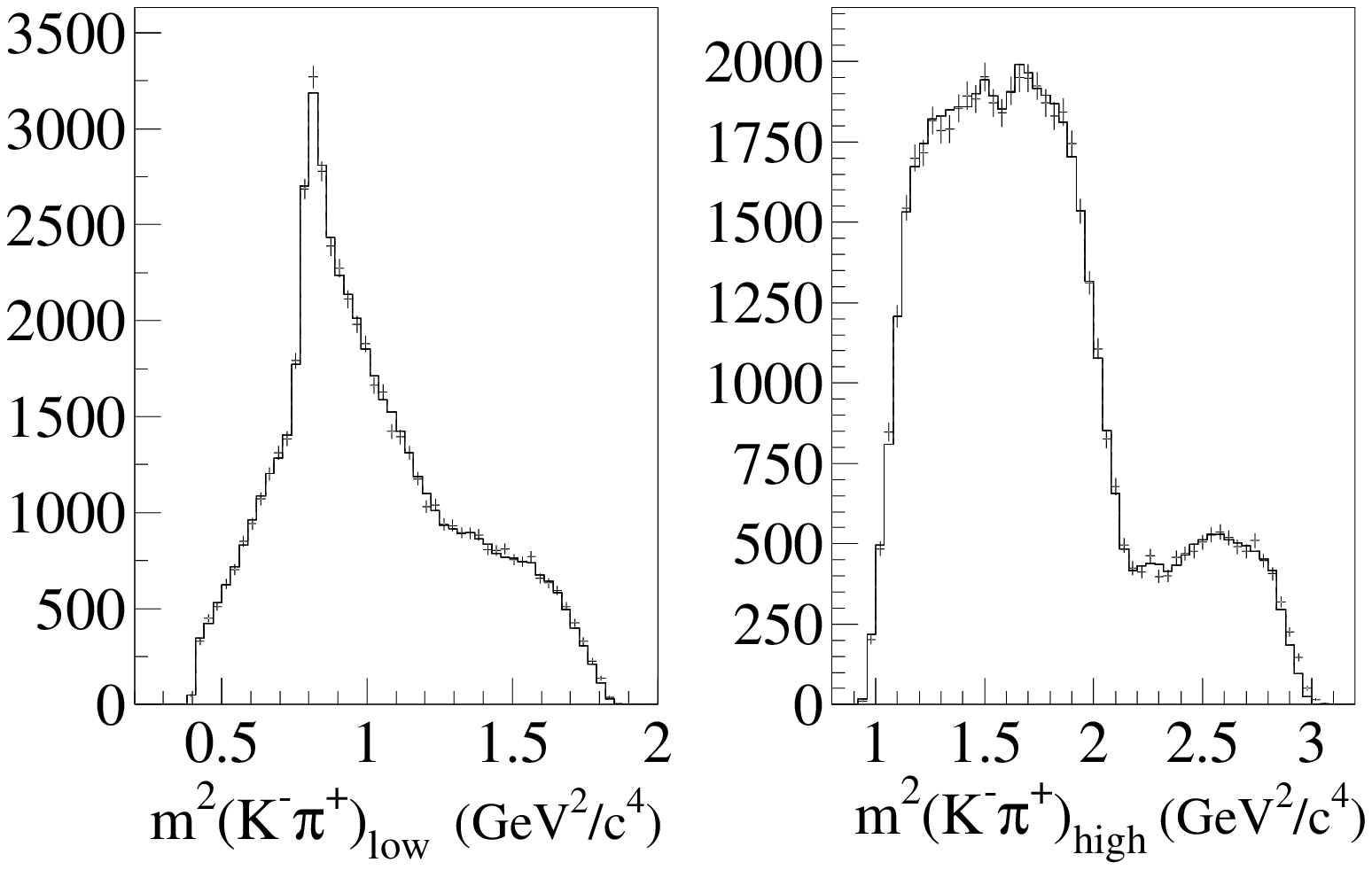}
\caption{Comparison between  the $K^{-}\pi^{+}\pi^{+}$ Dalitz plot projections and the MIPWA fit, 
with the lowest (left plot) and highest (right plot) $K^-\pi^+$ invariant mass squared.
In  the plots the solid histogram is a projection of the fit.}
\label{pwaproj}
\end{figure}

The Dalitz plot projections (highest and lowest $K^-\pi^+$ invariant mass squared) are plotted in 
Fig. \ref{pwaproj}, with the fit result superimposed (solid histograms).

\subsection{Systematic uncertainties}

Systematic uncertainties may come from different sources.
We have performed  {\em split sample} studies, in which the
data was divided into four sets of independent samples, according to the parent $D$ meson charge and momentum.
The \emph{split sample} component takes into account the possible systematics
introduced by a residual difference between data and Monte Carlo, due to a
possible mismatch in the reproduction of the $D^{+}$ production. A technique, employed in FOCUS and modeled 
after the \emph{S-factor method} from the Particle Data Group~\cite{pdg}, was used 
to try to separate true systematic variations from statistical fluctuations.
We found a small effect from the split sample studies.

A second class of studies is the {\em fit variant}, in which the fit of the whole data set
is performed under different conditions. Fit variants included changes in the background level and 
in the first derivatives of the spline at the edges of the $K^-\pi^+$ spectrum. The  {\em fit variant}
component can be estimated by the {\it r.m.s.} of the measurements.

The third and dominant source of systematic errors comes from the uncertainty in the parameters of the
P- and D-waves. This includes uncertainties on the values of the parameters  $r_R$ and $r_D$.
We repeated the fit changing by $\pm 1\sigma$, one at a time, the values of the mass and width of the 
high-mass vector resonances, according to the PDG, and of the parameters  $r_R$ and $r_D$. This component
is also estimated by the {\it r.m.s.} of the measurements.

The contributions of each source are quoted individually in Table \ref{pwa1}, \ref{pwa2}  and \ref{swavet}.
The overall systematic uncertainty was obtained adding in quadrature the three
components described above, and corresponds to the values in parentheses in Table \ref{swavet}.

\section{Summary and conclusions}

A Dalitz plot fit was performed with the MIPWA technique. The $K^-\pi^+$ S-wave 
amplitude was determined directly from data, with no assumption about its nature.
The only hypotheses are that the decay amplitude can be described by a sum of partial waves, and that
the P- and D- waves are well described by a coherent sum of Breit-Wigner amplitudes.

The MIPWA decay fractions are in good agreement with our previous analysis and with the E791 results.
A large dominance of the S-wave component is observed in this decay. 

The phase of the S-wave amplitude grows continuously across the elastic range, with a total variation 
of approximately 200$^o$. At the $K^-\pi^+$ threshold there is a phase difference of approximately 
-140$^o$ between the S- and P-waves. 

The phase variation of the S-wave measured in this analysis and that of E791 agree well, specially in 
the elastic range.
Our definition of the S-wave amplitude, eq. \ref{c0}, differs from that of E791. The
latter includes a Gaussian form factor, so one should compare the S-wave magnitude 
from our analysis to the product of the E791 Gaussian form factors and magnitude. We also find
a qualitative agreement between the S-wave magnitude measured by the two experiments.

FOCUS has performed a comprehensive study of the $D^+ \rightarrow K^-\pi^+\pi^+$  Dalitz plot. Using the 
same events, fits with the isobar model, the K-matrix formalism and the MIPWA were performed. The
three fits have equivalent goodness-of-fit. The decay fractions from all fits are in good agreement. 
In the isobar model there is a strong correlation between the nonresonant and $\kappa \pi$ modes. Although 
a good fit with this model is achieved, it is difficult to disentangle the contribution of these two
modes.

\begin{figure}
\includegraphics[width=10 cm]{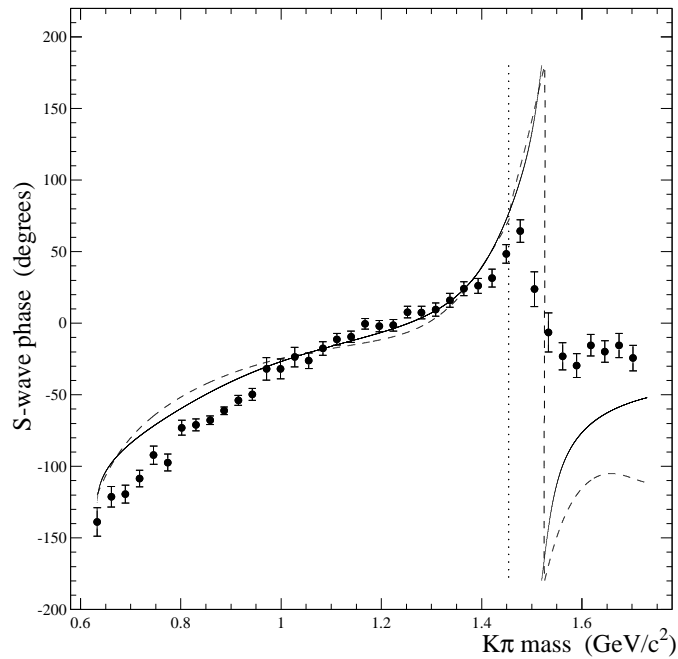}
\caption{Comparison between  the S-wave phase from the three different FOCUS fits of the $K^{-}\pi^{+}\pi^{+}$ 
Dalitz plot. 
Points with error bars are the result of the MIPWA fit. The solid line is the central value of the
isobar fit. The dashed line is the result of the K-matrix fit.}
\label{pwakmat}
\end{figure}

In Fig. \ref{pwakmat} the S-wave phase from the three fits are compared. All fits show a good agreement
in the interval 1~$< m_{K\pi} <$~1.35~GeV/c$^2$. The MIPWA phase is lower than those from the isobar/K-matrix
fits for $m_{K\pi} <$~1~GeV/c$^2$. In the high mass region the rapid variation of the phase is more pronounced
in the isobar/K-matrix fits than in the MIPWA.

\begin{figure}
\includegraphics[width=10 cm]{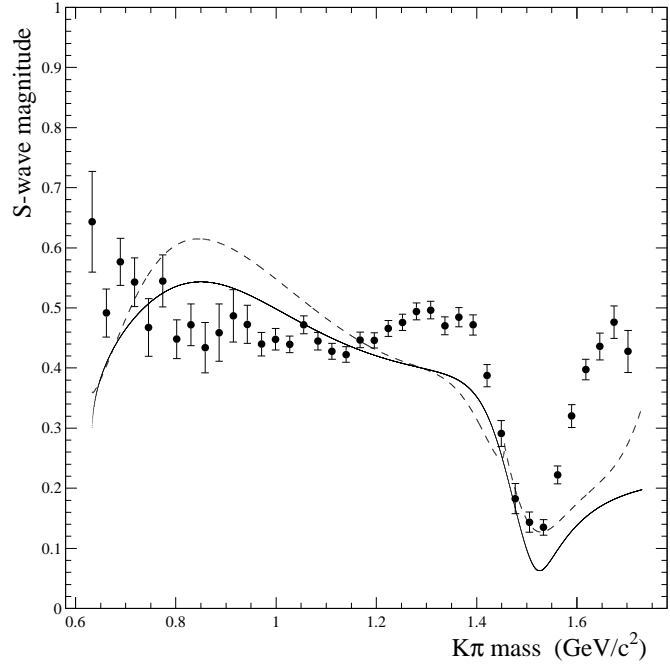}
\caption{Comparison between the S-wave magnitude from the three different FOCUS fits of the $K^{-}\pi^{+}\pi^{+}$ 
Dalitz plot. 
Points with error bars are the result of the MIPWA fit. The solid line is the central value of the
isobar fit. The dashed line is the result of the K-matrix fit.}
\label{pwakmat2}
\end{figure}

The S-wave magnitude from the three fits are compared in Fig. \ref{pwakmat2}. In the isobar and K-matrix
fits there is a broad maximum at around 0.9 GeV/c$^2$, which is absent in the MIPWA fit. In the region
1.2 $< m_{K\pi} <$ 1.4 GeV/c$^2$ the MIPWA magnitude has a bump whereas in the isobar and K-matrix the magnitude
decreases. In the high mass region, after the minimum, the magnitude from the MIPWA fit has a steeper variation 
than that of the isobar and K-matrix.

The $D^+ \rightarrow K^-\pi^+\pi^+$ decay offers an opportunity to access the $K^-\pi^+$ S-wave 
amplitude near threshold. Except for heavy flavor decays, no new data on the $K^-\pi^+$ system are foreseen. 
The ultimate goal is to extract the $I$=1/2  $K^-\pi^+$ elastic amplitude, 
where all resonances are contained. The result of the MIPWA fit, however, may include other effects, 
such as a possible contribution of the  $I$=3/2  amplitude, or an energy dependent phase introduced by 
three-body final state interactions. The road from the MIPWA S-wave to the  $I$=1/2  $K^-\pi^+$ elastic 
amplitude is, unfortunately, not direct. Input from theory is necessary. At this level of
statistics we are already limited by systematics, which are dominated by the uncertainties on 
resonance parameters.

%
%

\vspace{1.cm}

We wish to acknowledge the assistance of the staffs of Fermi National
Accelerator Laboratory, the INFN of Italy, and the physics departments of
the collaborating institutions. This research was supported in part by the
U.~S. National Science Foundation, the U.~S. Department of Energy, the
Italian Istituto Nazionale di Fisica Nucleare and Ministero della Istruzione
Universit\`a e Ricerca, the Brazilian Conselho Nacional de Desenvolvimento
Cient\'{\i}fico e Tecnol\'ogico and FAPERJ, CONACyT-M\'exico, and the Korea Research
Foundation of the Korean Ministry of Education.

%
%


\begin{thebibliography}{99}


\bibitem{cg} E.M. Aitala {\em et al.} (E791 Collaboration), 
Phys. Rev. Lett. {\bf 89}, 121801 (2002).

\bibitem{bes} M.~Ablikim \textit{et al.} (BES Collaboration), 
Phys. Lett. ~B{\bf598}, 149 (2004).

\bibitem{kmat} J.M.~Link \textit{et al.} (FOCUS Collaboration), 
Phys. Lett. ~B{\bf653}, 1 (2007).

\bibitem{CLEO-c} G.~Bonvicini \textit{et al.} (CLEO-c Collaboration), 
Phys.\ Rev.\ D{\bf 78}, 052001 (2008).

\bibitem{dg}  S. ~Descotes-Genon and B. ~Moussallam, 
Eur. Phys. J. {\bf C48}, 553 (2006).

\bibitem{babar} B.~Aubert \textit{et al.} (BaBar Collaboration), 
Phys.\ Rev.\ D{\bf 76}, 011102 (2007).

\bibitem{belle} D.~Epifanov \textit{et al.} (Belle Collaboration), 
arXiv:0706.2231.
 
\bibitem{lass} D. ~Aston \textit{et al.} (LASS Collaboration), 
Nucl. Phys. {\bf B296}, 493 (1988).

\bibitem{3pi} J.M.~Link \textit{et al.} (FOCUS Collaboration), 
Phys. Lett. ~B{\bf585}, 200 (2004).


\bibitem{bm} E.M.~Aitala \textit{et al.}, (E791 Collaboration), 
Phys. Rev. ~D{\bf73}, 032004 (2006).

\bibitem{photon} P.L.~Frabetti \textit{et al.}, (E687 Collaboration),
Nucl.~Instrum.~Meth. A329 (1993) 62.

\bibitem{WJohns}  J.M.~Link \textit{et al.}, (FOCUS Collaboration),
Nucl.~Instrum.~Meth. A516(2004) 364.

\bibitem{spectro}  P.L.~Frabetti \textit{et al.}, (E687 Collaboration),
Nucl.~Instrum.~Meth. A320 (1992) 519.


\bibitem{cerenkov} J.M.~Link \textit{et al.}, (FOCUS Collaboration),
Nucl.~Instrum.~Meth. A484 (2002) 270.

\bibitem{blatt} J.M. ~Blatt and V.F. ~Weisskopf, {\em Theoretical Nuclear Physics}
(John Wiley \& Sons, New York, 1952).

\bibitem{pdg} C.~Amsler \textit{et al.} (Particle Data Group), 
Phys. Lett. ~B{\bf 667}, 1 (2008).

\end{thebibliography}
\end{document}